\documentclass[12pt]{cernart}
\usepackage{graphics,epsfig,amssymb}

\def\piz     {\ensuremath{{\mathrm \pi}^{\circ}}}
\def\pip     {\ensuremath{{\mathrm \pi}^{+}}}
\def\Kp     {\ensuremath{{K}^{+}}}
\def\Kz     {\ensuremath{{K}^{\circ}}}

\def\Ke3{\ensuremath{\mathrm K_{e3}}}
\def\numtonue{\ensuremath{{\mathrm\nu}_{\mathrm\mu}\rightarrow{\mathrm\nu}_{\mathrm e}}}
\def\num {\ensuremath{\mathrm{\nu}_{\mathrm\mu}}}
\def\nue {\ensuremath{\mathrm{\nu}_{\mathrm e}}}

\def\Pnumtonue{\ensuremath{\mathrm{P}(\numtonue)}}
\def\dm2{\ensuremath{\Delta\mathrm{m}^2_{\mathrm{e}\mu}}}
\def\sin2tme{\ensuremath{sin^2 2\theta_{e\mu}}}
\def\cm{\ensuremath{\mathrm{cm}}}
\def\mm{\ensuremath{\mathrm{mm}}}
\def\GeVc{\ensuremath{\mathrm{GeV}/\mathrm{c}}}
\def\GeV{\ensuremath{\mathrm{GeV}}}
\def\MC{Monte Carlo}

\begin{document}
\begin{titlepage}
\docnum{CERN--PPE/96-181}
\date{December 5, 1996}
\title{CONCEPTUAL STUDY OF AN ``ANTI-TAGGED'' EXPERIMENT SEARCHING
FOR \numtonue\ OSCILLATION}
\begin{Authlist}
Lucio~Ludovici\Iref{a}, Piero~Zucchelli
\Instfoot{i1}{CERN, CH-1211 Geneva 23, Switzerland }
\end{Authlist}
 
\begin{abstract}
We study the conceptual feasibility of a high energy,
``short baseline'', zero background experiment to search for \numtonue\ 
oscillations and fully covering the area where the LSND experiment claims evidence.
The natural \nue\ background of the \num\ beam, from $K$ and $\mu$ decays in the 
decay tunnel, is suppressed by a hadron blind 
detector that vetoes, by time coincidence,
a possible \nue\ signal in the neutrino detector ({\sl anti-tagging technique}).
We discuss this new idea and we study a possible implementation in the old neutrino line of
the PS accelerator, which  at CERN offers the ideal $L/E$ ratio.
In the anti-tagged \num\ beam, the \nue\ contamination can be 
reduced by more than two orders of magnitude over conventional beams, 
down to $\nue/\num = 5\cdot 10^{-5}$.  
In an ideal appearance experiment using a $300\; t$ detector one would
expect after two years  $112$ events according to the LSND result,
with a background of $1.1\div 2.4$ events.  In case of a negative search,
the $90\%$ C.L. upper limit in the mixing angle 
would be $\sin2tme < 1.8\cdot 10^{-4}$ for large \dm2\ and $\dm2 <3.3\cdot 10^{-2}\; eV^2$
for maximal mixing.
\end{abstract}
%\vspace{2cm} 
\submitted{(Submitted to Nucl. Instr. and Methods A)}
\Instfoot{a}{On leave from INFN, Sezione di Roma ``La Sapienza'', Rome, Italy}
\end{titlepage}
 
%%%%%%%%%%
\section{Physics Motivations}
%%%%%%%%%%

In the current experimental scenario of the neutrino oscillation search,
claims of positive results come from
the solar \nue\ deficit \cite{kam,gal,sage,chlo}, 
the atmospheric anomalous $\nu_\mu/\nu_e$ ratio \cite{kam2,imb}, 
and the LSND $\bar{\nue}$ excess \cite{lsnd}. 
Whether the solar neutrino problem should be attributed to neutrino
oscillations \cite{bahcall,smirnov} or to the solar 
properties \cite{dar,confo,morrison,derujula} is still an open question.
On the atmospheric
neutrino anomaly, different results \cite{nusex,frejus,soudan2}
and interpretations \cite{gaisser,david}
%%\cite{soudan2,nusex,frejus,confo,david}
suggest a conservative attitude 
in drawing definite conclusions.
SuperKamiokande \cite{kam2kek} will provide soon an order of magnitude increase of
the data available to study the atmospheric neutrino deficit and, with 
SNO \cite{SNO} and
BOREXINO \cite{BOREXINO}, a deeper insight into the solar neutrino problem. 
On the other hand, no present or approved experiment is going to encompass
the area indicated by LSND for
possible \numtonue\ oscillations.
In the coming years, LSND will continue taking data and
the KARMEN experiment, after an upgrade to provide a better shield for
cosmic rays\cite{zeitniz},
should reach a sensitivity comparable to that of LSND.
It should be noted however that both
experiments have a similar neutron signature in the same energy range and rely on
background subtraction from similar sources.
Furthermore, the LSND present analysis of the data \cite{lsnd2} shows a limited potential 
in measuring the \dm2\ parameter.
\par
The variety of theoretical models \cite{smirnov,bilenky,ellis,wolfenstein,chou}
 predicting incompatible values
of neutrino masses and mixing angles and the lack of a
compelling
%coherent
experimental indication,  strengthen the belief that future experiments
should be mostly motivated by
 
\begin{itemize}

\item a {\sl discovery potential} which should
allow to confirm or disprove existing claims of evidence for oscillation with a
larger significance, possibly with a different detection
technique and experimental signature;
 
\item an {\sl experimental sensitivity} which in case of a negative result
should allow to exclude a large and still unexplored area of at least one or two
orders of magnitude in the $\Delta m^2$ and $sin^22\theta$ plane. 
\end{itemize}
\par
The subject of this paper is the feasibility of a high energy beam 
($E_\nu\gtrsim 1\; \GeV$), short baseline
appearance \numtonue\ experiment with ``zero background'' and maximum sensitivity for
$\dm2 \approx \mathrm{few} \; eV^2$.  This offers a unique opportunity to probe the
region where LSND claims evidence for oscillation with a different signature, ultimately
measuring the \dm2\ parameter in case of positive evidence.
\par
The main
problem to face using high energy \num\ beams consists in the irreducible background
due to \nue\ contamination produced by kaon and muon decays in the decay tunnel.
In conventional neutrino beams and detector setups \cite{bebc,ps191,charm,ccfr},
the expected \nue\ events due to the beam contamination
 are of the order of $0.5 - 2 \%$ with respect to the
\num\ interactions and this makes it very difficult to probe the oscillation
probability $P( \numtonue\ )=(0.31^{+0.11}_{-0.10} \pm 0.05)\%$ given by LSND.

%%%%%%%%%%
\section{Motivation for a zero background experiment}
%%%%%%%%%%
The search for \numtonue\ oscillation in the appearance mode in a 
conventional \num\ beam
requires an accurate modelling of the small \nue\ content in order
to subtract the corresponding background.
In the regime
where a {\it large background} from \nue\ contamination has to be subtracted,
the $90\%$ C.L. limit put on the
oscillation probability by a
negative search in an appearance experiment is 
ideally\footnote{All other background contributions are neglected.}
 given by
\begin{equation}
          \Pnumtonue \simeq 1.28 \sqrt{\frac{C_e}{N}
                  \left( 1+{(\Delta  C_e)^2\over C_e}N  + C_e \right) }
\label{eq:nuelimit}
\end{equation}
where $N $ is the observed number of neutrino interactions,
$C_e$ is the \nue\ contamination of the beam and  
$\Delta  C_e$ is its absolute uncertainty.
\par
In a {\it zero background} experiment, which could be defined by the condition
$C_e N \lesssim 1$, the limit on the oscillation probability
is given by
\begin{equation}
           \Pnumtonue = {2.3 \over N } \gtrsim 2.3 \cdot C_e
\label{eq:zeroblimit}
\end{equation}
\par
To give a numerical example, in a conventional beam with a \nue\ contamination
$C_e = 1\%$ and 
$\Delta  C_e/C_e = 3\%$\footnote{Usually the beam contamination uncertainty is about 10\%. A better
knowledge requires a dedicated experiment for the $K/\pi$ ratio measurement and tight control of the systematics
in the \MC\ simulation of the beam optics.}, as long as the
{\it zero background} condition is valid, the limit improves like
$1/N$. Then, below $\Pnumtonue \approx 2.3 \cdot C_e = 2.3\times 10^{-2}$, the limit
improves only as $1/\sqrt{N} $ up to the asymptotic limit
$\Pnumtonue \approx 1.28 \cdot \Delta  C_e = 3.8 \times 10^{-4}$, which is reached for
 \begin{equation}
            N  \gtrsim \frac{ (1+C_e)C_e } { {\Delta C_e}^2 }
%%{1\over C_e (\Delta  C_e/C_e )^2}
\label{eq:nmuasint}
 \end{equation}
In the example above this would correspond to $N \gtrsim 1.1\times 10^{5}$.
\par
In other terms, an experiment taking
$N  = 100,000$ events in such a beam could set a limit
$\Pnumtonue = 5.6\times 10^{-4}$
while the same limit could be obtained with $N  = 4100$ events
by a {\it zero background} experiment. This experiment would require a reduction of
the \nue\ contamination to $C_e\lesssim 1/N  = 2.4\times 10^{-4}$.
An ideal neutrino detector in a {\it zero background} beam would then reach the same limit set
by a detector $\sim 25$ times larger in a beam with the \nue\ contamination stated
above.
\par
From the point of view of studying the properties of a possible signal (like the
oscillation parameter \dm2 ) a {\it zero background} experiment offers the advantage of a
clear {\it event by event} identification of the signal, while in presence of 
background any information about the signal has to be extracted statistically 
from the candidate events.

%%%%%%%%%%
\section{Anti-tagging Principle}
%%%%%%%%%%
The principle  of the anti-tagging consists in a delayed time
coincidence between the  \nue\  
production time in the meson decay
and its interaction time in the neutrino detector.
With respect to the existing idea of a
{\sl tagged neutrino beam} \cite{tnf,bernstein},
the flavour identification is restricted to the \nue\  background 
events and the neutrino detector can be at any distance from the
source because no spatial correlation between the decay and the interaction
is required.
\par
\begin{figure}
\centering\epsfig{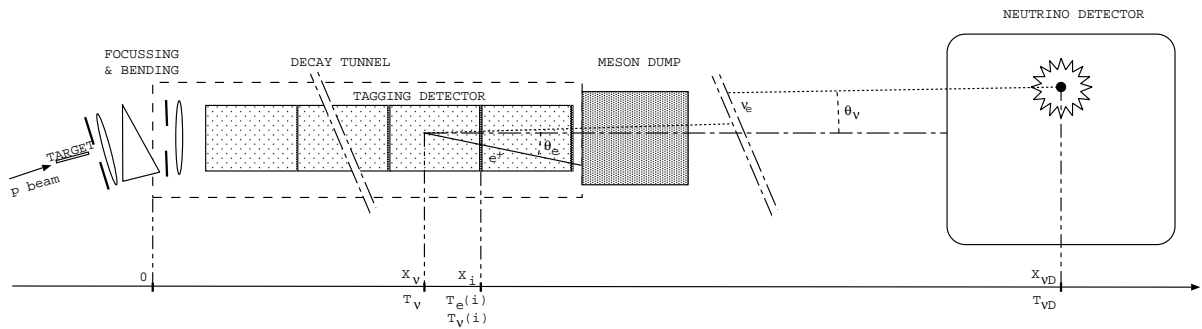}
\caption{Conceptual layout of the experiment.}
\label{fig:conclay}
\end{figure}

The aim of the anti-tagging is to suppress the \nue\ contamination
installing in the decay tunnel a detector capable to identify the production of each
\nue , thus vetoing the interactions which occur in the neutrino detector.
Indeed, the tagging detector measures the positron accompanying
the neutrino in
the decay $\mathrm{M}\rightarrow\nue \mathrm{e}^+ \mathrm{X}$.
In figure~\ref{fig:conclay} is depicted the schematic arrangement of the
experiment. Thin, planar modules installed in the decay tunnel
perpendicularly to the meson beam, detect the passage of
the positron measuring its crossing time $T_e(i)$ on the i-th module.
% with a resolution $\delta_t$.
A downstream detector records all neutrino events
and measures their interaction time $T_{\nu D}$.
\par
%, taking into account the neutrino time
%of flight in the approximation of a neutrino velocity $\beta_\nu=1$.

Given a neutrino interaction, the information of all tagging
modules is recorded  in order to look for possible positron signals
at time
 \begin{equation}
    T_e(i)  = T_{\nu D} - {(X_{\nu D}-X_i)\over c } + \Delta T(i)
 \label{eq:nudtime}
 \end{equation}
where $X_{\nu D}$ and $X_i$ are respectively the positions of the neutrino
interaction vertex and of the i-th tagging module.  The term 
$\Delta T(i)$ would be zero if the neutrino crossing on the 
i-th tagging module would be isochronous with the positron and if
$cos\theta_\nu=1$.  In general $\Delta T(i)$ depends on the beam energy 
and the detector geometry and it is given by
 \begin{equation}
\Delta T(i)= 
  {(X_{\nu D}-X_i)\over c} \left( {1-\frac{1}{cos\theta_\nu}} \right) + 
  {(X_i-X_\nu)\over c}\left[{1\over \beta_e cos\theta_e}
 - {1\over cos\theta_\nu}\right]
\label{eq:dtt}
 \end{equation}
where $X_\nu$ is neutrino production vertex position.
$\Delta T(i)$ can be estimated on average, or even evaluated event by event 
when the positron direction $cos\theta_e$ is measured as
soon as the positron is detected by more than one module.
\par
The residual uncertainty on $\Delta T(i)$ has to be smaller than the 
resolution $\delta t$  of the time anticoincidence between
$T_{\nu D}$ and $T_e(i)$ used to veto neutrino events.

\par
Another condition which has to be satisfied by the time resolution
$\delta t$ of the anticoincidence, concerns the random veto
due to accidental coincidence between an oscillation event 
and an uncorrelated positron:
the time resolution should be such that 
$\delta t\cdot f_t\ll 1 $, where $f_t$ is the tagging rate. 
\par
Each tagging module operates by detecting the Cherenkov light produced by the 
positrons
in the gas filling the decay tunnel, when the positron velocity $\beta_e$
exceeds the Cherenkov threshold $1/n$, where $n$ is the gas
refractive index.
The Cherenkov properties of different gases which
could be suitable for our purpose will be discussed later in detail,
but here we note that a gas radiator is needed in order to keep hadrons and muons below
the Cherenkov threshold.
\par
The Cherenkov photons produced along the positron path are emitted
in the forward direction with an aperture angle
$cos \Theta_\gamma=1/(\beta_e\cdot n)$.
They all reach the tagging module almost
simultaneously, filling a circular area around the positron impact point
({\sl Cherenkov spot})
with a constant {\sl radial} density 
%prortional to $1\over \mathrm{r}$, 
and within a radius
\begin{equation}
            r = d \cdot tan\Theta_\gamma
\label{eq:radius}
\end{equation} 
where d is the radiator length.
\par
It can be shown that, in the limit $tan\theta_e\ll 1/(n-1)$, the following
relation holds for the difference between
the positron crossing time
and the arrival time of the Cherenkov photons emitted at a distance $D_\gamma$
from the tagging module
 \begin{equation}
      T_\gamma - T_e \approx {D_\gamma\over c\beta_e }[n^2-1]
 \label{eq:gammadelay}
 \end{equation}
This delay of the Cherenkov photons with respect to the positrons is negligible
in every practical case, compared to the neutrino-positron timing.
 
%%%%%%%%%%
\section{Experiment design}
%%%%%%%%%%
% layout with numbers
\subsection{General layout}
In this section we present a possible implementation of the
anti-tagging idea, 
to prove its feasibility 
%%showing a possible solution, 
by addressing
the issues of tagging rate, 
efficiency, neutrino flux and background.
The optimization of the design and eventually
the identification of alternative or complementary options is left to a 
more comprehensive study. 
All the numbers quoted in the following sections
are based on the reference layout described below, and when different 
solutions are considered it will be explicitly stated.
\par
The study has been performed with a full simulation
of the target \cite{geant} and the magnetic optics \cite{turtle} 
to calculate the characteristics of the secondary meson beam.
The physics in the decay tunnel has been described with the JETSET
\MC\ \cite{jetset}, taking into account only decays and not
particle interactions.
\par
%%The proton beam extracted from the PS accelerator is transported to the
%%target using the TT7 extraction line.  
The secondary meson beam is focussed and
bent by $15^\circ$ with a magnetic system which transports the positive charge particles
into the decay tunnel.  The decay tunnel
%%, where the neutrinos are produced, 
is $80\,\mathrm{m}$ long and is instrumented with the tagging detector, which is
followed by a conventional dump to absorb all particles except neutrinos.  
\par
The tagging detector consists of $25$ tagging modules positioned along the tunnel.
Each module is a Cherenkov threshold detector consisting of a $3\,\mathrm{m}$ long gas 
radiator followed by a planar photon detector. The gas radiator is operated sligthly 
above the atmospheric pressure for gas purity considerations. 
The radiator and the photon detector 
are contained in a cylindrical vessel of $1 \,\mathrm{m} $ radius with thin windows
on the front and rear side.  The rear window is just on the back
of the photon detector. We estimate the thickness of materials traversed by the 
particles
to be less than $5\cdot 10^{-3}\,X_\circ$ per tagging module. 
\par
The neutrino detector is located 
%in the BEBC hall 
$810\,\mathrm{m}$ from the
center of the decay region.  For acceptance calculations, we assume a detector transverse square
section of $4\times 4\,\mathrm{m}^2$.    
%%%%%%%%%%
\subsection{Neutrino Beam}
%%%%%%%%%%
\subsubsection {Proton beam}
A neutrino beam energy of a few \GeV\ is suitable for a short baseline experiment with a
maximum sensitivity in the range $\dm2 \approx \mathrm{few} \; eV^2$.  
\par
In order to keep the tagging rate to an acceptable level, the
available proton intensity should be extracted onto the target as slowly
as possible. The optimal solution would be the accumulation in a storage
ring with a continuous extraction. Without accumulation,
 the anti-tagging is feasible provided a slow extraction scheme is adopted.
\par
At CERN the slow extraction is used both for the CPS and SPS proton accelerators.
Both facilities could be exploited taking advantage of the existing decay tunnels and 
experimental halls. The flux estimate presented in this study is based on the assumption 
of a primary proton beam from the CPS accelerator with the characteristics summarized below. 
\par
We assume a spill length of $500\; ms$ at a proton energy of $19.2\; GeV$.
The slow extraction is less efficient 
than the usual fast extraction, and the requirement of minimizing the proton
losses in the machine limits at present the intensity to $10^{13}$  p/cycle.
In the foreseeable future, after the LEP shutdown and before
the LHC era, the CPS proton availability could probably allow to allocate
up to 4 cycles of $2.4\; s$ each, in a supercycle of $14.4\; s$.
Then a neutrino experiment in an anti-tagged beam could run at the 
CPS with
an intensity of $4\cdot10^{13}$ protons every 14.4 seconds, with a
proton on target rate of $2\cdot 10^{13}\;s^{-1}$ during extraction.
%%This estimate is based on actual performances of the present slow extraction,
%%and probably a dedicated study could optimize intensity and rate.
\par
Based on the CPS performances in the last four years, we assume a running time of
$5600$ hours per year devoted to physics, with an accelerator efficiency of $90\%$.
In a two years data taking about $1.0\cdot 10^{20}$ protons on target could be expected. 
 
\subsubsection {Target and meson beam}

The secondary particles yield produced by $19.2 \; \GeVc$ protons  
impinging on a beryllium
target is simulated using GEANT \cite{geant}.  
The target is a cylindrical rod, parallel to the beam, 
with a diameter of $3\, \mm $ and a length of 
$110 \, \cm $ (corresponding to $2.7$ absorption lengths and $3.1$
radiation lengths).
\begin{figure}
\centering\epsfig{file=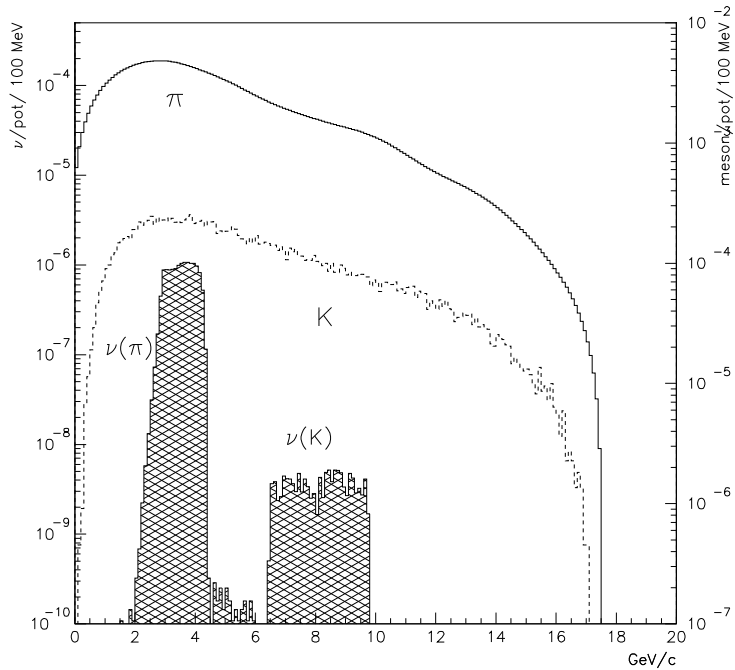,width=10cm}
\caption{Meson from target (right axis) and neutrino spectra in the detector (left axis). Both are
         normalized to the proton on target flux.}
\label{fig:spectra}
\end{figure} 
\par
Figure~\ref{fig:spectra} shows the spectra of \pip and \Kp\ produced in the target.
A traditional horn scheme for the focussing system
is incompatible with the slow extraction because it has to be operated
in short pulses \cite{horn}. 
A magnetic system consisting of quadrupoles and dipoles
can both focus and bend the meson beam from the target into the decay tunnel.
In addition to the charge selection, the bending removes from the meson 
beam the \Kz\ component, which is the main 
source of \nue\ background, and the direct photon yield
from the target.   The momentum acceptance of the focussing system cuts the low
momentum part of the secondary beam,  strongly 
suppressing the rate in the tagging detector due to the
positrons produced by soft kaon decays
and those produced in the target.
A main advantage in such a focussing scheme consists in the low \nue\
 background in the neutrino detector. 
The relative \nue\ flux
is about $0.1\%$ for a corresponding \num\ yield
of $1.42\times10^{-5}\; \num /pot$ on the neutrino detector.
The \num\ spectrum is shown in figure \ref{fig:spectra}.
\par
 The  meson beam focussing is not a
critical issue, because at this energy the neutrino beam divergence is
determined by the large
neutrino decay angle with respect to the parent meson
($24\,\mathrm{mrad}$  for $\pi_{\mu 2}$ and $64\,\mathrm{mrad}$ for 
$K_{\mu 2}$ on average). 
The meson beam divergence should be small enough 
to contain the secondary beam inside the tagging detector.
\par
In our simulation we assume a magnetic focussing system with an 
angular acceptance of $50\,\mu\mathrm{Sr}$, a momentum selection of 
$\Delta P/P = 20\%$ centered around  $P_0 =8.5\; \GeVc\ $ and a meson 
beam divergence of
 $3\,\mathrm{mrad}$ with a beam width of $10\; cm$.
\par  
As an example we have simulated \cite{turtle}
a possible scheme
where the target is followed by a collimator slit, a quadrupolar triplet,
a dipole and finally a quadrupolar doublet for the end focussing. 
The compactness of the end focussing section after the bending is an important 
feature because  the  presence of the magnets complicates the tagging of the 
meson decays.
A pure quadrupolar magnetic line, with non coaxial elements, could be an alternative 
solution for the
focussing system still allowing a bending angle \cite{jaapnoname}.  
An important difference with
respect to the previous option, where the bending is obtained using a dipole,  
is the absence of the charge selection of the mesons.
\par
The flux of minimum
ionizing particles in the central
region of a tagging module, where the beam intensity is maximum, 
is estimated to be $180\; \mathrm{MHz}/cm^2$, including also the secondary particles from 
decays in the tunnel. 
%%%%%%%%%%
\subsection{Tagging Detector}
%%%%%%%%%%
\subsubsection{Cherenkov light for tagging}
Cherenkov threshold gas detectors have been used in high energy physics
since a long time.
The properties of a few gases at STP are reported in table~\ref{table:gas}.
\begin{table}[htb]
\vspace{0.3cm}
\begin{center}
\begin{tabular} {|l||l|l|l|l|} 
\hline
 & $He$ & $Ne$ & $H_2$ & $CF_4$\\
\hline
$n-1\; (10^{-4} \mathrm{units} )$ & 0.35 & 0.67 & 1.38 & 4.0  \\
$\Theta _\gamma\; (\mathrm{mrad})$  & 8.37 & 11.5 & 22.6 & 28.2  \\
$\gamma_{thr}$ & 120 & 86 &43 & 35 \\
$I\; (eV)$ & 24.6 & 21.6 & 15.4 & 12.  \\
$e   \; E_{thr} \; (GeV)$ & 0.061 & 0.044 & 0.022 & 0.018 \\
$\mu \; E_{thr} \; (GeV)$ & 12.7 & 9.1  & 4.5  & 3.7 \\
$\pi \; E_{thr} \; (GeV)$ & 16.7 & 12.0 & 6.0  & 4.9  \\
$K  \; E_{thr} \; (GeV)$  & 59.2 & 42.4 & 21.2 & 17.3  \\
$p   \; E_{thr} \; (GeV)$ & 112.6& 80.7 & 40.4 & 32.9  \\
\hline
\end {tabular}\vspace{0.3cm}
\end {center}
\caption{Cherenkov properties of some gases}
\label{table:gas}
\end{table}
Helium and neon are natural candidates because at
atmospheric pressure all positrons
are above
threshold, while the Cherenkov emission due to hadrons and muons  
can be neglected after the momentum selection. 
The differential Cherenkov light yield is given by
\begin{equation}
 \frac{dN^\gamma}{dEdx} = \frac{\alpha Z^2}{\hbar c} 
             \left( 1 - \frac{1}{\beta^2 n(E)^2} \right)
\label{eq:yield}
\end{equation}
and is intrinsically small for light  gases
\footnote{For helium, in the visible spectrum, about 0.027 photons/cm 
are emitted at saturation.}.  The integrated light yield
increases with the bandwidth and then there is a clear advantage in
detecting light up to the extended ultraviolet region (EUV),
defined by a photon energy $E\lesssim 25\; eV$ ($\lambda \gtrsim 51.2\; nm$).  
Light noble gases are particularly
suitable because of their high ionization potential which determines the upper
frequency for the light transmission.
Appropriate EUV photon detectors have to face the problem that most materials
(in particular all solids) are not transparent.  Solutions
have been recently proposed and tested by 
\cite{charpa,charpb,charpc}.
\par
\begin{figure}
\centering\epsfig{file=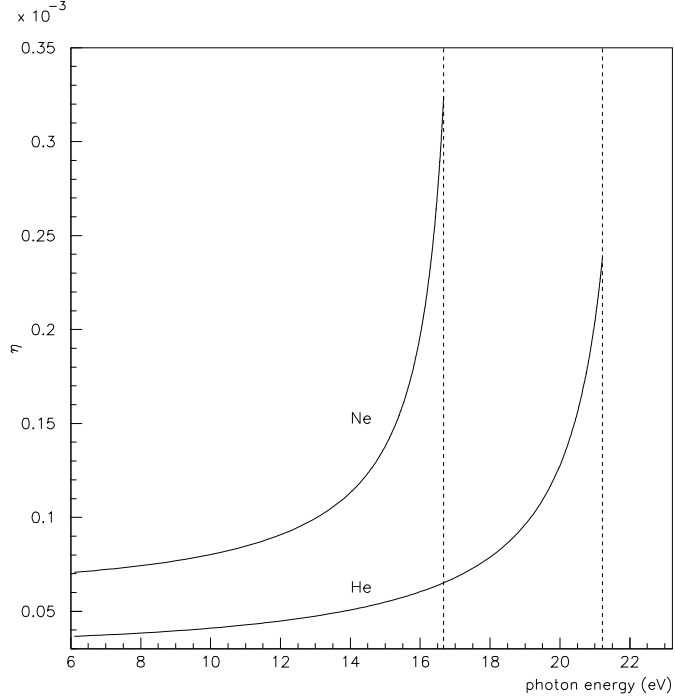,width=10cm}
\caption{$\eta=n-1$ as a function of the Cherenkov photon energy.}
\label{fig:eta}
\end{figure}
\begin{figure}
\centering\epsfig{file=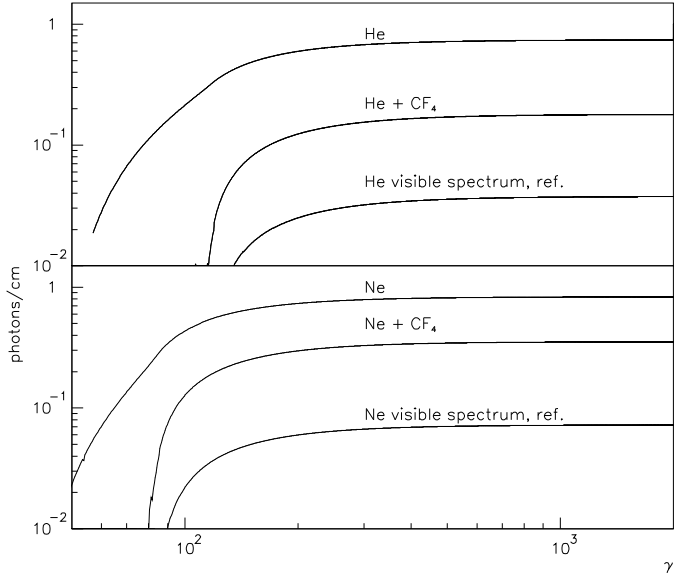,width=10cm}
\caption{Cherenkov light yield as a function of $\gamma=E/mc^2$ for 
         He and Ne. The effect of $CF_4$ traces is shown together
         with the light yield in the visible ($350\div 500$ nm).}
\label{fig:querenchov}
\end{figure}

In the optical region the refractive index is essentially constant 
but in the EUV, close to
the allowed dipole transitions, the variation of the refractive 
index has to be taken
into account in evaluating the integrated light yield from equation (\ref{eq:yield}). 
$\eta=n-1$ can be calculated as a function of the photon energy
 by extrapolation of 
measurements in the optical and UV region \cite{optic1}.
The result is shown in figure~\ref{fig:eta}  in the energy range $6\div 21.6$ eV for helium
and $6\div 16.7$ eV for neon. 
The integrated $\mathrm{photon/cm}$ yield and the Cherenkov threshold for a 
few gas radiators are shown in figure~\ref{fig:querenchov}.
The addition of a proper 
``Cherenkov quencher'', i.e. traces of gas with a lower ionization potential, 
allows to
tune the Cherenkov threshold and the light yield. 
The understanding, and the way to control such mechanism, 
can be studied
in a test beam where 
the measurement of the light yield  allows
to determine experimentally, for the first time, the helium refractive
index in the EUV region.

\subsubsection{Photon detector structure}
 
The characteristics of a suitable photon detector can be summarized as follows:
\begin{itemize} 
\item UV-EUV photon detection;
\item time resolution $\lesssim 1 \; ns$;
\item spatial granularity ($1 \div 5\; cm$ or less);
\item low radiation length ($\leq 0.01 X_0$);
\item high rate capability;
\item radiation hardness.
\end{itemize}
\par 
The required time resolution is larger than the uncertainty on 
the $\Delta T(i)$ term defined in equation \ref{eq:dtt}, which is
about $90\; ps$ on average.
\par 
In this section we show a possible detector based on the
Micro-Gap Chamber (MGC) technology \cite{mgc1}. 
Other possible solutions could be studied, such as 
MICROMEGAS \cite{giom} or MCP 
based EUV detectors, as well as standard detection techniques for visible
light coupled to fast wavelength shifters.

\par
\begin{figure}
\centering\epsfig{file=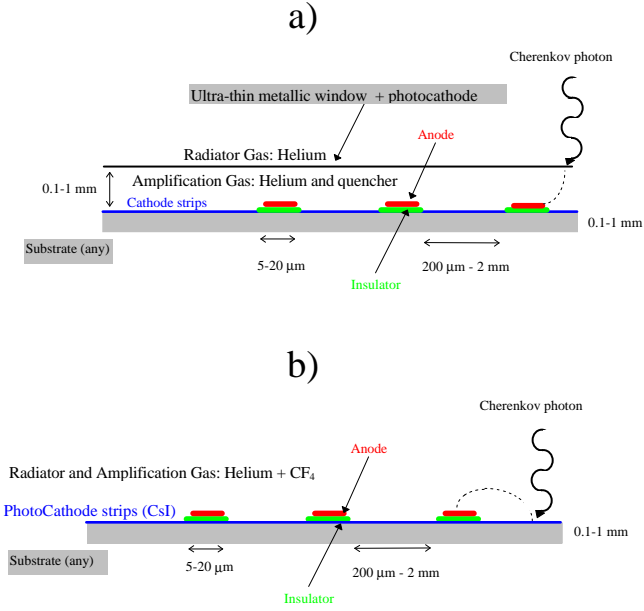,width=10cm}
\caption{Two possible options for the photon detector with MGC: 
         (a) with a thin metallic window and (b) windowless.}
\label{fig:mgcsetup}
\end{figure}

The development of EUV photon detectors is particularly interesting
because it could result in generic hadron blind Cherenkov
detectors with fast timing and short radiator
length, appealing for many applications in high energy physics.
The MGC technology has already been  applied to
Cherenkov photon detection in the visible range \cite{bozzo}.
Similar solutions could be adapted to detect photons in the 
EUV region (figure~\ref{fig:mgcsetup}).
\par
A first option consists in using {\sl metallic window MGC}. 
The quartz window that
delimites the drift space in UV sensitive MGC could be replaced by 
an ultra-thin window supported by a thicker micromesh 
that acts as gas separator between the drift volume and the radiator, as
suggested in \cite{metallic}. The ultra-thin window (about $50\; nm$)
could be a metallic film with a CsI photocathode deposited by sputtering or
the CsI photocathode itself. The drift distance
can be $500\; \mu m$ or less, in order to increase the hadron blindness.
The window between the drift space and the radiator 
allows
to choose the quencher and the radiator gas mixture
independently,
so that the MGC can operate at the maximum gain
with the most efficient radiator.
\par
In this scheme, a
positron signal can give up to 
$100\; p.e. \times 5 \cdot 10^4 =5 \cdot 10^6$ 
electrons collected within a Cherenkov spot 
of $2.5\;cm$ radius ($3.5\; cm$ for neon). 
Despite the strong suppression due to the small drift volume and the use of low Z gases,  electrons can be produced by minimum ionizing particles both on the
photocathode surface and, with a lower amplification, in the drift volume.   
% The signals from minimum ionizing particles are strongly suppressed due to 
% the small drift volume and the use of low-Z gases.  Electrons can be
% produced through ionization both in the photocathode surface and
In the central region of the
tagging module, where the beam intensity is higher, 
the average signal from
minimum ionizing particles on an area corresponding to a Cherenkov spot
can be estimated to be of the order of $1\div 2$ 
photoelectrons.  Then the fake
positrons induced by the  minimum ionizing particles can be efficiently
reduced by applying a pulse height threshold.
\par
A second option consists in 
{\sl windowless MGC}, where the radiator gas is the same mixture
used for the amplification of the electron cascade. The photoelectrons are
 produced directly on the strip cathodes (see figure \ref{fig:mgcsetup}) 
sputtered by a thin CsI
film, as already studied in many high energy physics applications \cite{nappi}. 
CsI photocathodes in light noble gas atmosphere, when an extraction field is applied,
allow very high quantum efficiencies (up to $50\%$) in the EUV energy range \cite{RD26}.
With respect to the previous configuration, the amplification occurs in a single
step and is consequently smaller (about $5\cdot 10^3$).
In this configuration the quencher fills the radiator volume and then quenchers 
with high ionization potential should be preferred in order to have a higher light yield.
The light transmission and the
gas multiplication should be optimized by maximization of the overall signal,
using for example
$CF_4$ ($I \simeq 12 \; eV$) or DME ($I \simeq 10 \; eV$) as quenchers.
\par
We estimate that  up to
$30 \; p.e. \times 5\cdot 10^3=1.5\cdot 10^5$ electrons ($50\; p.e.$ for Ne)
are collected for an average positron. 
The main advantage of this option, despite of the smaller signal amplitude, 
is the reduced sensitivity to minimum ionizing particles, which is only due
to backward emission of secondary electrons from the photocatode surface.  
\par
For the above considerations on the sensitivity to minimum ionizing particles 
and on the possibility to tune the Cherenkov threshold, the tagging rate is 
dominated
by the positrons produced in the decay tunnel, and amounts to
$f_t\approx 98\; \mathrm{MHz}$. Since a positron traverses on average $6.1$
modules, only $8.2\;\mathrm{MHz}$ are due to singles and the
rest to coincidences of two or more modules. 
\par
The numbers given should be confirmed by experimental data, addressing
also the issues of radiation hardness, sensitivity to 
minimum ionizing particles and to
scintillation light, as well as the actual performances of a prototype
tagging module in a beam environment.
 
%%%%%%%%%%
\subsection{Neutrino Detector Requirements}
%%%%%%%%%%
The detector should be designed
to identify a possible electron in the low multiplicity final state 
of a few \GeV\ neutrino interaction. 
We will not address in this paper the issues of the neutrino detector in
detail.  The general detector requirements are
\begin{itemize}
\item a mass of the order of $\sim 300\, t$;
\item a time resolution of $\sim 1~ns$ or better;
\item $e/\piz$ separation at a level of $\leq 10^{-3}$;
\item electron and muon identification;
\item cosmics background discrimination.
\end{itemize}
A good energy resolution on neutrino interactions is an important feature for
the \dm2\ measurement.
To cope with these requirements several detector options are possible using
technologies which are presently available or could be available with a
reasonable R\&D effort.
\par
 A liquid argon TPC imaging detector,  characterized
by an excellent separation $e/\piz=10^{-4}$ \cite{aladin}, 
 satisfies all the requirements with the exception of the timing.
This however could be provided by the detection of the
scintillation light, or even better the Cherenkov light, produced in the liquid
argon.
\par
A large volume water Cherenkov detector could also be a possible neutrino
detector with an intrinsically good timing.
\par
Another viable solution consists in a conventional sandwich
calorimeter, with thin
$(\lesssim 1/5 X_\circ)$ absorbers and trackers, 
where the timing could be provided
by the trackers themselves (for instance resistive
plate chambers \cite{RPC}) or by additional planes of fast scintillator.
A different approach could be a fully active liquid scintillator
calorimeter, segmented in a cell structure to 
provide a suitable tracking capability.
 
%%%%%%%%%%
\section{Experiment sensitivity}
%%%%%%%%%%
In this section 
%%we estimate the sensitivity of the conceptual experiment in the \numtonue\
%%search. 
we evaluate the \nue\ composition of the beam, 
 the anti-tagging efficiency, and the irreducible \nue\ contamination.
The possible sources of
background in the \nue\ detection
are discussed, in order to assess the total contamination in the
oscillation search.
We finally estimate the experiment sensitivity
for two years data taking, and the \dm2\ measurement potential in case
of a positive \numtonue\ signal detection.
 
% acceptances
\subsection{Anti-tagged beam background}
We identify the following sources of \nue\ background, of which the associated
positrons
are not detected by the anti-tagging detector:
\begin{itemize} 
\item \nue\ produced before the bending
optics, which reach the far detector;
\item decays in uninstrumented regions of the decay tunnel;
\item tagging modules acceptance.
%\item $\mu^+\rightarrow e^+ \nu_e \bar{\nu_\mu}$ decays outside
%the decay tunnel;
\end{itemize} 
\par 
The \nue\ from decays occurring before the bending
 have a  much
softer spectrum
than the \num\ from the decay tunnel. 
Applying a loose $1.5$
GeV cut on the neutrino energy,
the contribution on the detector can be estimated to be 
$\lesssim 0.2\cdot10^{-5} \nue/\num $.
\par
The beginning of the decay tunnel, where the focussing magnets are located, 
and the region close to the beam dump, are potential sources of 
untagged \nue .
Assuming that the photon detector of the first tagging module is 
positioned $6\; m$
from the beginning of the tunnel, and that the magnets 
occupy the first $3\; m$,
we estimate $42\%$ of the positrons produced in the magnet 
region cannot be detected,
and this accounts for a background of $2.7\cdot 10^{-5} \; \nue/\num$.
We assume that the \nue\ production in the last $50 \; cm$ of the tunnel
cannot be detected. Taking also into account the decays inside the beam dump, 
the contribution is evaluated to be $0.5\cdot 10^{-5}\; \nue/\num$.
Secondary muons from the meson decays are another source of \nue\ background,
which has to be estimated considering their penetration in the dump
and possible decay outside the instrumented region.
However, the \nue\ contribution coming from muons escaping the decay
tunnel has a soft spectrum and a broad angular distribution (due to the decay 
chain and the muon energy loss in matter) and from simulation it is
 negligible.
\par
The geometrical inefficiency due to large angle positrons escaping undetected from the tagging
modules accounts for $2.0\cdot 10^{-5}$ \nue/\num\ background, including both \Ke3\ and 
muon decays (respectively $89\%$ and $11\%$ of the \nue\ in the neutrino detector).
\par
We conclude that the 
irreducible background is about $5\cdot 10^{-5}$ \nue/\num,
which improves by more than two orders of magnitude the \nue\
contamination with respect to conventional neutrino beams.
 
\subsection{Background in the neutrino detection}
The main sources of background in the detection of \nue\ interactions are
\piz\ resonant and coherent production and $\nu_\mu e$ interactions. 
\par
The cross section of coherent production, measured in the interesting energy range by the 
Aachen-Padova collaboration \cite{bobi1} and Gargamelle \cite{gargpi}, 
can be estimated in about
$20\div 40 \cdot 10^{-40} cm^2/\mathrm{nucleus}$, depending on the nuclear
composition of the target and in agreement with the theoretical expectations \cite{rscoh}.  
\par
Neutral pions can be produced incoherently  in nucleon scattering, and fake
the electron signature. The Rein and 
Sehgal model \cite{rs} agrees well with Gargamelle \cite{garga2} data predicting
$\sigma=8.6\cdot 10^{-40} cm^2/\mathrm{nucleon}$.
\par
The  $\num e\rightarrow\num e $ properties are well known
from theory, and were measured by the CHARM II experiment.
The cross section is typically $1.6\cdot 10^{-42} E_\nu (GeV)\; cm^2$ \cite{charm2e}, 
and the characteristic kinematics
can be distinguished because of the forward electron signature:
$ E\theta^2<2 m_e$.
 This background 
 can be reduced below the foreseen sensitivity
by requiring, for the oscillation candidate events, 
 a minimum angle between the observed electron and
the beam axis.

\subsection{Oscillation sensitivity}
The \num\ 
flux onto the neutrino detector 
is $1.4 \cdot 10^{15}\; \num$ in two years of data taking. 
%%On an isoscalar target,
%% the quasi-elastic cross section, including
%% resonances production, has been evaluated to be
%%$0.85\cdot 10^{-38}cm^2/\mathrm{nucleon}$ \cite{qe}. 
The inclusive charged current cross section
has been measured in our energy range to be
$\sigma_{\nu N}=(2.45 \pm 0.15)\cdot 10^{-38} cm^2/ \mathrm{nucleon}$
at $<E_\nu>=3.6 \; GeV$ \cite{garga}.
This corresponds to about $36,500$ neutrino
interactions (corrected for accidental vetoes) in a $300\; t$ detector. 
From previous considerations on background, the irreducible contamination
due to the beam \nue\ component is $1.8$ events.
To evaluate the sensitivity in the oscillation search we restrict the sample  
to the range $2<E_\nu<5\; \GeV $, which loosely corresponds to neutrinos 
produced in the pion decays: in that case $36,200$ \num\ events are left, with a 
background of $1.0$ \nue\ events.
\par
The overall \piz\ contamination would be $0.14 \div 1.4$ events, depending on the separation
capability ($e/ \piz\ = 10^{-4} \div 10^{-3}$).
We have assumed, for simplicity of normalization, $100\;\%$ detection efficiency
both for signal and background, since this value strongly depends on the
neutrino detector and analysis strategy.
For the same reason we do not take into account that the \piz\ and the 
$\num e\rightarrow\num e $
backgrounds are
further reduced applying the mentioned cut on the reconstructed neutrino energy.
\par
\begin{figure}
\centering\epsfig{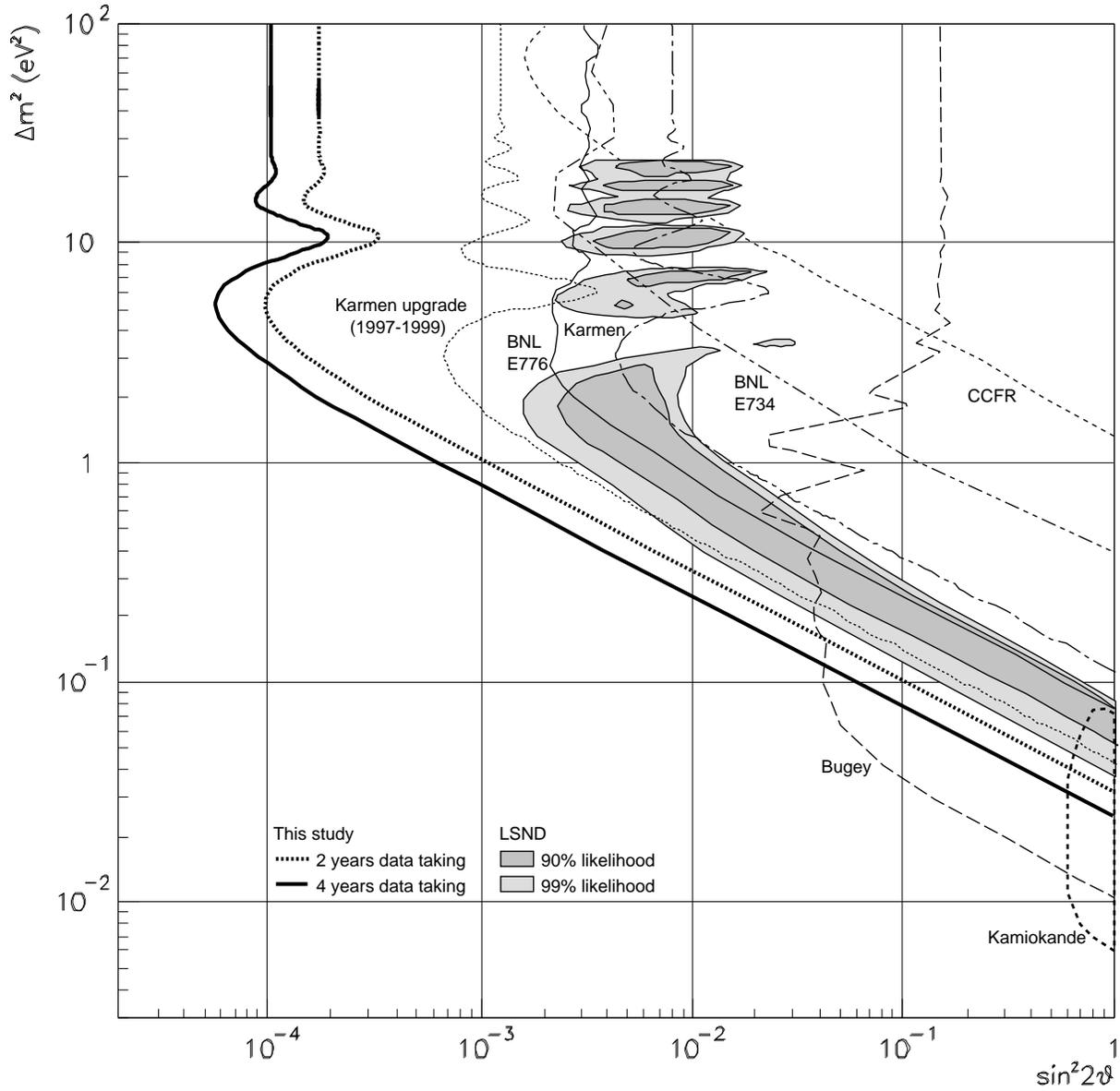}
\caption{Exclusion plots from a negative search in a two and four years data 
taking period. 
The LSND favored regions (90\% and 99\% likelihood probability) are shown 
as well as the KAMIOKANDE inclusion plot.  The present limits from BNL-E776,
KARMEN, BNL-E734, Bugey, CCFR are reported. For comparison is also shown the 
limit that would be expected by the upgraded KARMEN detector in a three years 
negative search.}
\label{fig:exclusion}
\end{figure}

All the relevant numbers are summarized in table \ref{table:summary}.
The exclusion plot resulting from a negative search, calculated following the procedure
described in ref.
\cite{helen}, is shown in 
figure \ref{fig:exclusion}. 
The LSND region is fully
covered, and part of the \numtonue\ atmospheric neutrino region is also probed.

\subsection{ \dm2\ measurement}
\begin{figure}
\centering\epsfig{file=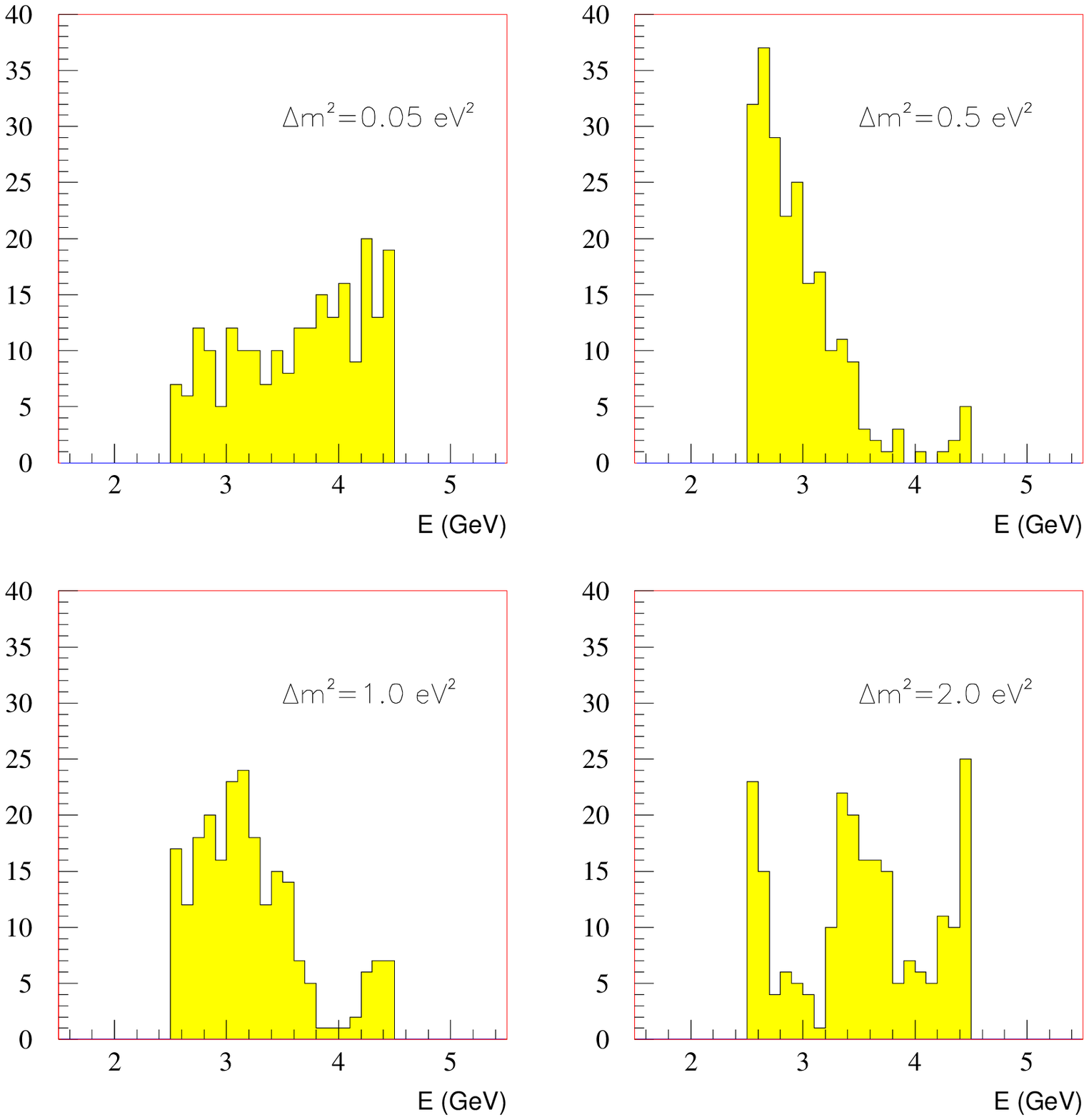,width=10cm}
\caption{Spectra of oscillation events for 
%%$\dm2=0.05,0.5,1.0,2.0 \; eV^2$
different values of \dm2. 
%For comparison is also shown 
%(dashed line) the spectrum of \nue\ events that would be expected
%from a conventional beam with a  
%$0.6\%$ \nue/\num\ contamination.
}
\label{fig:delta}
\end{figure}

If the LSND hypothesis on \numtonue\ oscillation is correct, the expected
signal after two years run is $112\pm 40$ events, with a background of 
$1.1 \div 2.4$ events. 
We can profit from the small uncertainty on the neutrino flight path
($\Delta L/L \approx 3\% \; RMS$)  
to measure \dm2\ from the energy distribution
of the oscillation events.
In figure 
\ref{fig:delta} are reported the 
energy distributions of the candidates for different values of \dm2,
 in the hypothesis of a detector resolution
%%$$\frac{\Delta E}{E} = \frac{5\%}{\sqrt{E(GeV)}}$$
$\Delta E/E = 5\%/\sqrt{E(GeV)}$. In the figure are shown 
224 oscillation events that could be collected with a two years 
``discovery extension'' of the run.

Independently from \sin2tme, \dm2\ can be measured below a few $eV^2$, while a lower limit on \dm2\ 
is set for higher values. 
In the last case, an extension run with a higher meson momentum selection or a close smaller detector
could increase the region where the \dm2\ modulation is measurable.

\begin{table}[htb]
\vspace{0.3cm}
\begin{center}
\begin{tabular} {|l|l|} 
\hline
%%characteristics &  \\
%%\hline \hline
Proton Energy & $19.2\; \GeV$ \\
Duration of data taking & 2 years \\
Integrated protons on target  & $1.0\cdot 10^{20}$ \\
Proton rate during extraction & $2\cdot 10^{13}\; \mathrm{Hz}$ \\
\num/pot on detector& $1.42\times 10^{-5}$ \\
\nue/\num (w/o anti-tagging)& 0.1\% \\
\nue/\num (with anti-tagging)& $5\cdot 10^{-5}$ \\
Tagging rate & $\approx98\; \mathrm{MHz}$ \\
Required $e/\piz$ separation& $10^{-4}\div 10^{-3}$ \\
$<E_\nu>$& $3.6\;\GeV$  \\
$<L>$& $810\; m$  \\
$\Delta L/L$ (RMS) & 3\%  \\
\num\ events ($\epsilon _{det}=1$ , $E_\nu=2\div 5\; \GeV$) & 36,200 \\
\nue\ background events ($\epsilon _{det}=1$ , $E_\nu=2\div 5\; \GeV$) & 1.0 \\
$\piz$ background events & $0.14\div 1.4$ \\
Expected \nue\ oscillation events from LSND claim& $112$  \\
\hline
\end {tabular}\vspace{0.3cm}
\end {center}
\caption{Main parameters and event rates.}
\label{table:summary}
\end{table}
%%%%%%%%%%
\section{Conclusions}
%%%%%%%%%%
We have presented the new idea of an anti-tagged \num\ beam, based on the time
coincidence between a positron detected in the meson decay tunnel and the 
neutrino interaction.  The meson decay tunnel is instrumented with noble gas 
Cherenkov detectors, which tag the positron produced in association
with the \nue .
\par
We have shown the conceptual feasibility of this idea discussing a possible 
implementation at the CERN PS.  A high energy \num\ beam with an effective \nue\ 
contamination of $5\cdot 10^{-5}$ can be achieved resurrecting the old PS neutrino
beam line.  This corresponds to a reduction of the \nue\ beam contamination
by more than two orders of magnitude  with respect to conventional neutrino beams.
\par
This anti-tagged \num\ beam can be exploited for a {\sl zero background} appearance
search for \numtonue\ oscillation, with a sensitivity which is an order of
magnitude better than any present or approved experiment.
\par
According to the LSND result, $112$ oscillation events could be collected in
a two years run with a background of $1.1\div 2.4$ events.
This sample could allow an analysis of the neutrino energy modulation
for an unmistakable 
oscillation signature and \dm2\ measurement.
%%%%%%%%%%
\section{Acknowledgments}
%%%%%%%%%%
It is a pleasure to warmly thank our colleagues from the CHORUS
collaboration for their encouragement. We acknowledge the comments and 
suggestions we received from many of them and from the participants to
the workshops on "Future Neutrino Experiments" at CERN.  
We are grateful to J.Boillot, R.Bellazzini, R.Cappi and D.J.Simon for
the valuable information received in informal discussions.
\bibliographystyle{unsrt}
\bibliography{biblio}

\begin{thebibliography}{10}

\bibitem{kam}
K.S.~Hirata et~al.
\newblock {\em Phys. Rev. Lett.}, 66:9, 1991.

\bibitem{gal}
W.~Hampel et~al.
\newblock {\em DAPNIA-SPP-96-10}, July 1996.
\newblock Subm. to, Phys. Lett., B.

\bibitem{sage}
J.~N.~Abdurashitov et~al.
\newblock {\em Nucl. Phys. B (Proc. Suppl.)}, 38:60, 1995.

\bibitem{chlo}
B.T.~Cleveland et~al.
\newblock {\em Nucl. Phys. B (Proc. Suppl.)}, 38:47, 1995.

\bibitem{kam2}
Y.~Fukuda et~al.
\newblock {\em Phys. Lett. B.}, 335:237, 1994.

\bibitem{imb}
R.~Becker-Szendy et~al.
\newblock {\em Phys. Rev. Lett.}, 69:1010--1013, 1992.

\bibitem{lsnd}
C.~Athanassopoulos et~al.
\newblock {\em Phys. Rev. C}, 54:2685, November 1996.
\newblock LA-UR-96-1582.

\bibitem{bahcall}
J.~Bahcall et~al.
\newblock {\em astro-ph/9610250}, June 1996.
\newblock 17th International Conf. on Neutrino Physics and Astrophysics -
  NEUTRINO '96, Helsinki, Finland, 13-20 Jun 1996.

\bibitem{smirnov}
A.~J. Smirnov.
\newblock {\em hep-ph/9511239}, November 1995.

\bibitem{dar}
A.~Dar.
\newblock {\em astro-ph/9611014}, November 1996.
\newblock 17th International Conf. on Neutrino Physics and Astrophysics -
  NEUTRINO '96, Helsinki, Finland, 13-20 Jun 1996.

\bibitem{confo}
G.~Conforto et~al.
\newblock {\em hep-ph/9606226}, June 1996.
\newblock UB FI 96-1, To be publ. in Astr. Phys.

\bibitem{morrison}
D.R.O. Morrison.
\newblock {\em Nucl. Phys. B, Proc. Suppl.}, 48:567, 1996.

\bibitem{derujula}
A.~De Rujula and S.L. Glashow.
\newblock {\em hep-ph/9208223}, August 1992.
\newblock HUTP-92-A038.

\bibitem{nusex}
M.~Aglietta et~al.
\newblock {\em Europhys. Lett.}, 8:611, 1989.

\bibitem{frejus}
K.~Daum et~al.
\newblock {\em WUB 95-03}, February 1995.
\newblock Subm. to, Z. Phys. C.

\bibitem{soudan2}
W.~W. M.~Allison et~al.
\newblock {\em PDK-570}, November 1996.
\newblock Subm. to, Phys. Lett.

\bibitem{gaisser}
T.K. Gaisser.
\newblock {\em hep-ph/9611301}, November 1996.
\newblock BA-96-50 and references therein.

\bibitem{david}
D.~Saltzberg.
\newblock {\em Phys. Lett. B}, 355:499, April 1995.

\bibitem{kam2kek}
Y.~Suzuki.
\newblock {\em presented at Int. Conf. on HEP, Warsaw}, July 1996.
\newblock to be publ. in Proceedings.

\bibitem{SNO}
D.~Sinclair et~al.
\newblock {\em Nuovo Cimento C}, 9:308--317, 1986.

\bibitem{BOREXINO}
G.~Ranucci et~al.
\newblock {\em Nucl. Instr. and Meth. A}, 315:229--235, 1992.

\bibitem{zeitniz}
K.~Eitel et~al.
\newblock {\em Proc. of the 8th Rencontres de Blois: Neutrinos, Dark Matter and
  the Universe}, June 1996.
\newblock To be published.

\bibitem{lsnd2}
C.~Athanassopoulos et~al.
\newblock {\em LA-UR-96-1326}, May 1996.

\bibitem{bilenky}
S.~M.~Bilenky et~al.
\newblock {\em Phys. Rev. D}, 54:1881, 1996.

\bibitem{ellis}
J.~Ellis et~al.
\newblock {\em Phys. Lett. B}, 292:189--194, 1992.

\bibitem{wolfenstein}
L.~Wolfenstein.
\newblock {\em hep-ph/9506352}, June 1995.
\newblock CMU-HEP-9505.

\bibitem{chou}
K.C. Chou and Y.L. Wu.
\newblock {\em hep-ph/9610300}, October 1996.
\newblock DOE-ER-01545-675.

\bibitem{bebc}
C.~Angelini et~al.
\newblock {\em Phys. Lett. B}, 179:307, 1986.

\bibitem{ps191}
G.~Bernardi et~al.
\newblock {\em Phys. Lett. B}, 181:173, November 1986.

\bibitem{charm}
F.~Bergsma et~al.
\newblock {\em Phys. Lett. B}, 157:469, 1985.

\bibitem{ccfr}
A.~Romosan et~al.
\newblock {\em hep-ex/9611013}, November 1996.
\newblock NEVIS-1529.

\bibitem{tnf}
V.~V.~Ammonosov et~al.
\newblock {\em Proposal SERP-E-152}.

\bibitem{bernstein}
R.~H.~Bernstein et~al.
\newblock {\em Fermilab P-788}.

\bibitem{geant}
Application~Software Group.
\newblock {\em GEANT Detector Description and tools}, June 1993.
\newblock CERN program library long writeups Q123, version 3.21.

\bibitem{turtle}
D.~C. Carey.
\newblock {\em SLAC 246}, March 1982.

\bibitem{jetset}
T.~Sjostrand.
\newblock {\em Computer Physics Communications}, 39:347, 1986.

\bibitem{horn}
S.~Van der Meer.
\newblock {\em CERN 61-07}, February 1961.

\bibitem{jaapnoname}
J.~Panman.
\newblock {\em CERN 83-02}, II:146, February,10 1983.
\newblock Yellow Report.

\bibitem{charpa}
Y.~Giomataris and G.~Charpak.
\newblock {\em Nucl. Inst. Meth. A}, 310:589, 1991.

\bibitem{charpb}
Y.~Giomataris et~al.
\newblock {\em Nucl. Inst. Meth. A}, 323:431, 1992.

\bibitem{charpc}
M.~Chen et~al.
\newblock {\em Nucl. Inst. Meth. A}, 346:120, 1994.

\bibitem{optic1}
P.W. Langhoff and M.~Karplus.
\newblock {\em J. Opt. Soc. Am.}, 59:863, July 1969.

\bibitem{mgc1}
F.~Angelini et~al.
\newblock {\em Nucl. Instr. and Meth.}, 335:69, October 1993.

\bibitem{giom}
Y.~Giomataris et~al.
\newblock {\em DAPNIA/SED 95-04}, December 1995.
\newblock Submitted to Nucl. Instr. and Meth.

\bibitem{bozzo}
F.~Angelini et~al.
\newblock {\em INFN PI/AE 95/03}, May 1995.
\newblock Submitted to Nucl. Instr. and Meth.

\bibitem{metallic}
A.~Braem et~al.
\newblock {\em CERN AT/94-33 (ET)}, September 1994.

\bibitem{nappi}
J.~Seguinot et~al.
\newblock {\em Nucl. Inst. Meth. A}, 297:133, 1990.

\bibitem{RD26}
A.~Breskin et~al.
\newblock {\em Nucl. Inst. Meth. A}, 344:163, 1994.

\bibitem{aladin}
M.~Bonesini et~al.
\newblock {\em SPSLC/I205}, June 1995.
\newblock CERN/SPSLC 95-37.

\bibitem{RPC}
R.~Santonico and R.~Cardarelli.
\newblock {\em Nucl. Inst. Meth. A}, 187:377, 1981.

\bibitem{bobi1}
H.~Faissner et~al.
\newblock {\em Phys. Lett. B}, 125:230, May 1983.

\bibitem{gargpi}
E.~Isiksal et~al.
\newblock {\em Phys Rev. Lett.}, 52(13):1096, March 1984.

\bibitem{rscoh}
D.~Rein and L.M. Sehgal.
\newblock {\em Nucl. Phys. B}, 223:29, 1983.

\bibitem{rs}
D.~Rein and L.M. Sehgal.
\newblock {\em Annals of Physics}, 133:79, 1981.

\bibitem{garga2}
W.~Lerche et~al.
\newblock {\em Phys. Lett. B}, 135:45, September 1978.

\bibitem{charm2e}
P.~Vilain et~al.
\newblock {\em Phys. Lett. B}, 302:351, 1993.

\bibitem{garga}
S.~Ciampolillo et~al.
\newblock {\em Phys. Lett. B}, 84:281, June 1979.

\bibitem{helen}
O.~Helene.
\newblock {\em Nucl. Inst. Meth.}, 212:319, January 1983.

\end{thebibliography}
%\begin{thebibliography}{99}
%\end{thebibliography}
 
\newpage
 
\end{document}